\documentclass[runningheads]{llncs}
\usepackage[T1]{fontenc}
\usepackage{graphicx}
\usepackage{amsmath}
\usepackage{mathpartir}
\usepackage[misc,geometry]{ifsym}
\usepackage[symbol]{footmisc}
\usepackage[most]{tcolorbox}
\usepackage{color,soul,float}
\usepackage{xcolor}
\definecolor{mauve}{rgb}{0.58, 0, 0.82}
\usepackage{listings}
\usepackage{subcaption}
\usepackage{hyperref}
\usepackage{minted}
\setminted{fontsize=\footnotesize}

\definecolor{annotationbg}{HTML}{D0F0C0}
\definecolor{annotationfg}{HTML}{1B5E20}

\definecolor{declbg}{HTML}{E8F5E9}
\definecolor{declfg}{HTML}{1B5E20}

\definecolor{singlebg}{HTML}{FFF6D1}
\definecolor{singlefg}{HTML}{8A6D1F}

\definecolor{ctrlbg}{HTML}{FFE4DE}
\definecolor{ctrlfg}{HTML}{C62828}

\definecolor{condbg}{HTML}{F0E5FF}
\definecolor{condfg}{HTML}{6A1B9A}

\definecolor{iterbg}{HTML}{E0F7FA}
\definecolor{iterfg}{HTML}{006064}

\definecolor{blockbg}{HTML}{E1BEE7}
\definecolor{blockfg}{HTML}{311B92}

\newtcbox{\annotationbox}{
    on line,
    arc=0pt,
    colback=annotationbg,
    colframe=annotationfg!70!black,
    coltext=annotationfg!40!black,
    boxsep=1pt,
    left=3pt,
    right=3pt,
    top=1pt,
    bottom=1pt,
    height=15pt,
    valign=center,
    boxrule=0pt
}

\newtcbox{\declbox}{
    on line,
    arc=0pt,
    colback=declbg,
    colframe=declfg!70!black,
    coltext=declfg!40!black,
    boxsep=1pt,
    left=3pt,
    right=3pt,
    top=1pt,
    bottom=1pt,
    height=15pt,
    valign=center,
    boxrule=0pt
}

\newtcbox{\singlebox}{
    on line,
    arc=0pt,
    colback=singlebg,
    colframe=singlefg!70!black,
    coltext=singlefg!40!black,
    boxsep=1pt,
    left=3pt,
    right=3pt,
    top=1pt,
    bottom=1pt,
    height=15pt,
    valign=center,
    boxrule=0pt
}

\newtcbox{\ctrlbox}{
    on line,
    arc=0pt,
    colback=ctrlbg,
    colframe=ctrlfg!70!black,
    coltext=ctrlfg!40!black,
    boxsep=1pt,
    left=3pt,
    right=3pt,
    top=1pt,
    bottom=1pt,
    height=15pt,
    valign=center,
    boxrule=0pt
}

\newtcbox{\condbox}{
    on line,
    arc=0pt,
    colback=condbg,
    colframe=condfg!70!black,
    coltext=condfg!40!black,
    boxsep=1pt,
    left=3pt,
    right=3pt,
    top=1pt,
    bottom=1pt,
    height=15pt,
    valign=center,
    boxrule=0pt
}

\newtcbox{\iterbox}{
    on line,
    arc=0pt,
    colback=iterbg,
    colframe=iterfg!70!black,
    coltext=iterfg!40!black,
    boxsep=1pt,
    left=3pt,
    right=3pt,
    top=1pt,
    bottom=1pt,
    height=15pt,
    valign=center,
    boxrule=0pt
}

\newtcbox{\blockbox}{
    on line,
    arc=0pt,
    colback=blockbg,
    colframe=blockfg!70!black,
    coltext=blockfg!40!black,
    boxsep=1pt,
    left=3pt,
    right=3pt,
    top=1pt,
    bottom=1pt,
    height=15pt,
    valign=center,
    boxrule=0pt
}

\newtcolorbox{outcodebox}{
    enhanced,
    colback=white,
    colframe=black,
    boxrule=1pt,
    arc=0pt,
    outer arc=0pt,
    width=\linewidth,
    top=1pt,
    bottom=1pt,
    before skip=0pt,
    after skip=0pt
}

\newtcolorbox{innercodebox}{
    enhanced,
    colback=gray!10,
    colframe=gray!50,
    boxrule=1pt,
    arc=0pt,
    outer arc=0pt,
    width=0.95\linewidth,
    center,
    left=0pt,
    right=0pt,
    top=1pt,
    bottom=1pt,
    before skip=1pt,  
    after skip=1pt    
}

\definecolor{codegreen}{rgb}{0,0.6,0}  
\lstdefinelanguage{diff}{  
  morecomment=[f][\color{blue}]{@@},     
  morecomment=[f][\color{red}]-,         
  morecomment=[f][\color{codegreen}]+,       
  morecomment=[f][\color{red}]{---}, 
  morecomment=[f][\color{codegreen}]{+++},
}

\usepackage{color,appendix}
\usepackage{multirow}
\usepackage{multicol, cuted}
\usepackage{lineno}
\usepackage{enumitem}

\lstdefinelanguage{Coq}%
  {morekeywords={Variable,Inductive,CoInductive,Fixpoint,CoFixpoint,%
      Definition,Lemma,Theorem,Axiom,Goal,Save,Grammar,Syntax,Intro,%
      Trivial,Qed,Symmetry,Simpl,Rewrite,Apply,Elim,Assumption, Module, Type,%
      Left,Cut,Case,Auto,Unfold,Exact,Right,Hypothesis,Pattern, Include,%
      Constructor,Defined,Fix,Record,Proof,Induction,Hints,let,in,%
      Parameter,Split,Red,Reflexivity,Transitivity,Opaque,%
      Transparent,Inversion,Absurd,Generalize,Mutual,Cases,of,end,Analyze,%
      AutoRewrite,Functional,Scheme,params,Refine,using,Discriminate,Try,%
      Require,Ensure,Load,Import,Scope,Open,Section,End,match,Ltac,%
      Instance,Class,With,unsigned,long,%
      bind,as,Inv,__u32,%
      Let,int,char,NULL,SEC,void%
	},%
   sensitive, %
   morecomment=[n]{(*}{*)},%
   morestring=[d]",%
   literate=
   {>->}{{$\rightarrowtail$}}2{->}{{$\rightarrow$}}1
   {==>}{==>}1
   {=>}{{$\Rightarrow$}}1
   {<->}{{$\leftrightarrow$}}2
   {|--}{{$\vdash$}}1
   {exists}{$\exists$}1 {forallx}{$\forall$}1
   {odot}{$\odot$}1 {Odot}{$\bigodot$}1
   {otimes}{$\otimes$}1 {Otimes}{$\bigotimes$}1
   {oplus}{$\oplus$}1  {Oplus}{$\bigoplus$}1
   {inx}{$\in$}1
   {star}{{$\star$}}1
   {mapsto}{{$\mapsto$}}1
   {wedge}{{$\wedge$}}1
   {nvdash}{{$\nvdash$}}1
   {sim}{{$\sim$}}1
   {prec}{{$\prec$}}1
  }[keywords,comments,strings]%

\newcommand{\myalt}[0]{\;$|$\;}

\usepackage{tikz}
\usetikzlibrary{positioning, shapes.geometric}
\definecolor{blizzardblue}{rgb}{0.67, 0.9, 0.93}
\definecolor{pastelorange}{rgb}{1.0, 0.7, 0.28}
\definecolor{celadon}{rgb}{0.67, 0.88, 0.69}

\usepackage{caption}
\usepackage[inference,ligature,reserved,shorthand]{semantic}
\usepackage{amsmath, amsfonts}
\usepackage{mathtools}
\usepackage{threeparttable}
\usepackage{proof}
\usepackage{mathpartir}
\usepackage{centernot}

\newcommand{\field}[2]{\textbf{field\_addr}(#1, #2)}
\newcommand{\store}[2]{\textbf{data\_at}(#1, #2)}

\begin{document}
\title{QCP: A Practical Separation Logic-based C Program Verification Tool}
\titlerunning{QCP: A C Verification Tool}

%
\author{Xiwei Wu\inst{1} \and Yueyang Feng\inst{1,\footnotemark[1]} \and Xiaoyang Lu\inst{1, \footnotemark[1]} \and Tianchuan Lin\inst{1} \and Kan Liu\inst{1} \and Zhiyi Wang\inst{2} \and  Shushu Wu\inst{1} \and Lihan Xie\inst{1} \and Chengxi Yang\inst{1} \and Hongyi Zhong\inst{1} \and Zihan Zhang\inst{1} \and Juanru Li\inst{1} \and Naijun Zhan\inst{2} \and Zhenjiang Hu\inst{2} \and Qinxiang Cao\inst{1, \footnotemark[2]} }
\renewcommand{\thefootnote}{\fnsymbol{footnote}}
\footnotetext[1]{These authors contributed equally to this work.}
\footnotetext[2]{Corresponding Author}

\renewcommand{\thefootnote}{\arabic{footnote}}
\authorrunning{Wu. et al.}
\institute{Shanghai Jiao Tong University \and Peking University }
\maketitle  
\begin{abstract}
As software systems increase in size and complexity dramatically, ensuring their correctness, security, and reliability becomes an increasingly formidable challenge. Despite significant advancements in verification techniques and tools, their practical application to complex, real-world systems is often hindered by critical gaps in both automation and expressiveness. 
To address these difficulties, this paper presents \textbf{Qualified C Programming Verifier (QCP)}, a novel verification tool that integrates annotation-based automatic verification with interactive proving using Rocq. QCP employs symbolic execution and a separation logic entailment solver to automatically discharge many verification obligations, while deferring more complex obligations to Rocq for manual proof. Furthermore, QCP includes a VS Code extension designed to enhance proof efficiency and support a deeper understanding of both the program behavior and verification outcomes.

\end{abstract}
\section{Introduction}

Software verification tools have made significant advancements, providing robust frameworks to ensure program correctness. Existing verification tools can be primarily classified into three predominant categories: (1) fully automated systems (e.g. Infer~\cite{10.5555/1986308.1986345}) that focus on shape properties, such as memory safety and the absence of specific faults like null pointer dereferences, using built-in heuristics and algorithms for predefined predicates; (2) annotation-based systems (e.g., VeriFast~\cite{VeriFast}, Viper~\cite{Viper}, Hip/Sleek~\cite{hipexample} and Smallfoot~\cite{10.1007/11804192_6}) which verify not only shape properties but also some functional correctness properties by leveraging built-in SMT solvers and user-provided annotations; (3) interactive systems (e.g., VST~\cite{VST}, Iris~\cite{10.1145/3022670.2951943} using Rocq) capable of verifying complex functional correctness but requiring substantial manual effort to write proof code in proof assistants.

Ideally, a verification tool should achieve the following:
\begin{itemize}[leftmargin=4.5em, labelindent=1em]
    \item[(A)] Support the verification of complex functional correctness properties, comparable to interactive proof assistants.
    \item[(B)] Minimize human intervention to the greatest extent possible, similar to fully automated and annotation-based tools. For simple programs and safety properties, verification should be achievable solely by writing annotations, without requiring manual proof construction in a proof assistant.
    \item[(C)] Provide real-time feedback during development.
    For example, immediately following the input of a precondition and a line of code, the tool should run symbolic execution and show the results before further code is written. If the execution fails, the tool should highlight the problematic line and report the error.
\end{itemize}
Existing tools either only support (A) or (B); quite a few of them support (A) and (B) simultaneously,  let alone (C) (to the best of our knowledge, none of them support (C)). 
This paper introduces \textbf{Qualified C Programming (QCP)}, a C program verification tool that targets all three requirements above.

For (A) and (B), QCP adopts well-established design elements from existing tools. Specifically, (1) QCP handles the memory manipulation of the C language using separation logic, which introduces an additional connective logic separating conjunction ($*$).  $P * Q$  asserts that $P$ and $Q$ hold for the disjoint heap regions. (2) It allows users to annotate programs with assertions outlining a proof skeleton, enabling separation-logic-based symbolic execution to generate verification conditions automatically. (3) For C function calls, QCP requires users to provide function specifications that describe the functions' behavior. During symbolic execution, QCP verifies that the preconditions of callee functions are satisfied and derives the postconditions for the calling context. (4) To automate the process of checking verification conditions, users can employ QCP’s built-in SMT solver or add customized separation logic heuristics. (5) QCP allows users to manually write Rocq proof code to fill proof gaps, which are about user-defined predicates that cannot be verified automatically by an SMT solver.  Therefore, QCP can combine the advantages of annotation-based and interactive-based verifiers.

However, regarding requirement (C), conventional verification tools typically require users to provide a fully annotated program before initiating the verification process. This approach presupposes that developers possess thorough familiarity with both the program code and the tool's internal verification mechanisms—specifically, the ability to anticipate whether a given program can be automatically verified under the current assertions. In contrast, QCP introduces a Visual Studio Code extension and a web interface that visualize intermediate verification states and partial results during the verification process. This functionality provides users with insight into program behavior and the underlying verification process, enabling them to incrementally develop and refine annotations, which is especially useful when working with complex or unfamiliar code.

\paragraph{Outline.}
This paper makes two primary contributions: (1) a novel integration of annotation-based and interactive verification methodologies (Section~\ref{sec::overview}), (2) an incremental verification support via a Visual Studio Code (VS Code) extension or a web interface (Section~\ref{sec::VSCode}). Additionally, Section~\ref{sec::implementation} presents the implementation of QCP. Section~\ref{sec::eval} presents comprehensive evaluation results of QCP across diverse sample programs. Section~\ref{sec::related} discusses related work, including comparative analysis with existing verification tools to demonstrate QCP's efficacy and practical utility, while Section~\ref{sec::conclu} concludes with findings and suggests directions for future research.
\section{Overview of QCP} \label{sec::overview}

\begin{figure}[!htbp]
    \vspace{-2.0em}
    \centering
    \includegraphics[width=\textwidth]{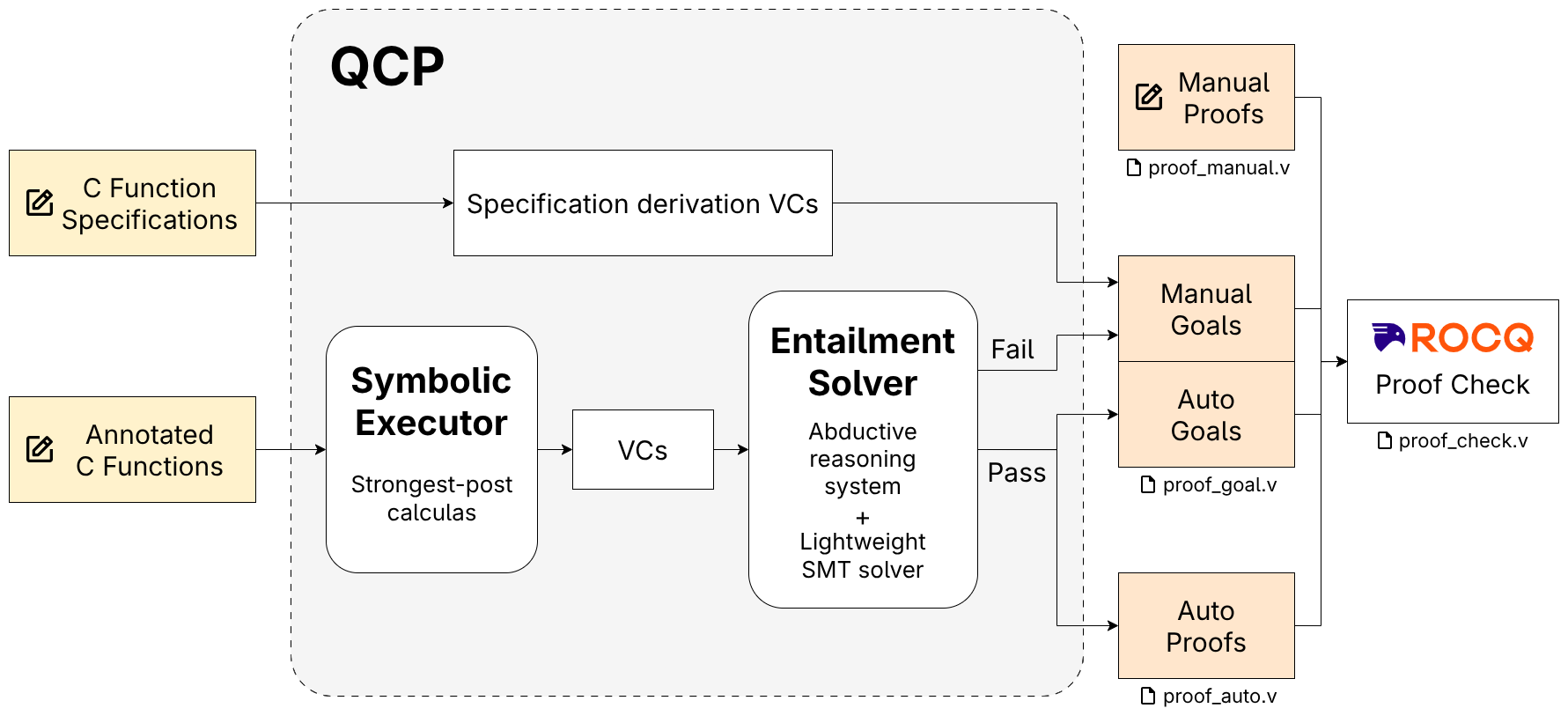}
    \vspace{-1.0em}
    \captionsetup{justification=raggedright, singlelinecheck=false}
    \caption{The components of QCP.}
    \vspace{-2.0em}
    \label{fig:QCPframe}
\end{figure}

QCP takes annotated C code as input and generates verification conditions. The straightforward conditions are discharged automatically, as shown in Figure 1, while the complex ones are left for manual proof within the expressive logic of the Rocq proof assistant.


\subsection{C module verification}

QCP supports C module verification. In other words, we can verify all functions defined in a C file against their specifications, assuming that any function merely declared in that file --- but implemented elsewhere --- satisfies its own specification. In such cases, QCP requires that each function be accompanied by at least one function specification. Figure~\ref{fig:sll_free} is a minimal example of C module verification in QCP: \lstinline{free_node_list} is only declared in this module and probably implemented elsewhere, and \lstinline{sll_free} is implemented in this module. Both functions are attached with their specifications. The predicate \lstinline{listrep}, conventionally used to describe the structural properties of singly-linked lists, is defined as follows:

\begin{lstlisting}
listrep(x) = x == NULL && emp || 
    exists v y, x != NULL && x -> next == y && x -> data == v && listrep(y)
\end{lstlisting}

The \lstinline{listrep} predicate specifically characterizes that pointer \lstinline{x} references a singly-linked list, without concerning itself with the list's contents or other attributes. Here, the expression \lstinline{x->next} indicates that we hold both the permission for \lstinline{x} and the access permission to its \lstinline{next} field. The syntax of QCP is designed to make sure that the .c file is still a legal C program that can pass typical compiler checks like gcc. Indeed, this file says, if \lstinline{free_node_list(x)} correctly frees the memory permission of \lstinline{x -> data} and \lstinline{x -> next}, then the implementation of \lstinline{sll_free(x)} is also correct w.r.t. its own specification, i.e., \lstinline{sll_free(x)} frees all elements in the linked list with head point \lstinline{x}.

\begin{figure}[htp]
  \vspace{-1.0em}
  \centering
\begin{minted}{c}
// The function free_list_node and the definition of struct list 
// are imported via the header file (.h).
struct list {
   int data;
   struct list *next;
};

void free_list_node(struct list * x)
/*@ With d n
    Require x -> data == d && x -> next == n 
    Ensure emp */;

void sll_free(struct list * x)
/*@ Require listrep(x)
    Ensure emp */
{
  /*@ Inv listrep(x) */
  while (x != NULL) {
    /*@ exists v n, x -> data == v && x -> next == n && listrep(n) */
    struct list *y = x -> next; 
    free_list_node(x); 
    x = y;
  }
}
\end{minted}
  \vspace{-1.0em}
  \captionsetup{justification=raggedright, singlelinecheck=false}
  \caption{example of \lstinline{sll_free}. The predicate \lstinline{listrep} describes the structural properties of singly-linked lists. }
  \vspace{-1.0em}
  \label{fig:sll_free}
\end{figure}

In practical software verification, it is often necessary to reason about functions at different levels of abstraction. A function may be equipped with both a concrete, implementation-oriented specification and more abstract specifications that capture its high-level behavior. To handle this, QCP mandates that each function be assigned one primitive function specification, which is used directly in verifying the function itself. Any other function specifications associated with it can be derived from this primitive specification and utilized in verifying functions that call it. QCP automatically generates verification conditions for function specification derivation, guaranteeing that all secondary specifications logically follow from the primary one, thereby maintaining consistency and completeness at the module level. 

\begin{figure}[!htp]
  \centering
  \vspace{-1.0em}
\begin{minted}{c}
typedef struct LOS_DL_LIST {
  int pstData;
  struct LOS_DL_LIST * pstPrev, * pstNext;
} LOS_DL_LIST;

void LOS_ListAdd(LOS_DL_LIST *list, LOS_DL_LIST *node)
/*@ primitive_spec
  With (x: Z) (a: Z)
  Require list -> pstNext == x && x -> pstPrev == list && 
      node -> pstData == a && 
      has_permission(&(node -> pstPrev)) * has_permission(&(node -> pstNext))
  Ensure  x -> pstPrev == node && node -> pstNext == x &&
      list -> pstNext == node && node -> pstPrev == list &&
      node -> pstData == a */;

void LOS_ListAdd(LOS_DL_LIST *list, LOS_DL_LIST *node)
/*@ head_insert_spec <= primitive_spec
  With (a: Z)
  Require node -> pstData == a && store_dll(list) * 
    has_permission(&(node -> pstPrev)) * has_permission(&(node -> pstNext))
  Ensure store_dll(list) */
{
  node->pstNext = list->pstNext;
  node->pstPrev = list;
  list->pstNext->pstPrev = node;
  list->pstNext = node;
}

void LOS_ListTailInsert(LOS_DL_LIST *list, LOS_DL_LIST *node)
/*@ With (a: Z)
  Require node -> pstData == a && store_dll(list) * 
    has_permission(&(node -> pstPrev)) * has_permission(&(node -> pstNext))
  Ensure store_dll(list)
*/
{  LOS_ListAdd(list->pstPrev,node); }
\end{minted}
  \vspace{-1.0em}
  \captionsetup{justification=raggedright, singlelinecheck=false}
  \caption{Specification of \lstinline{LOS_ListAdd} and \lstinline{LOS_ListTailInsert} from LiteOS. The predicate \lstinline{store_dll(x)} represents a circular doubly-linked list storage structure, where \lstinline{x} denotes the sentinel node.}
  \vspace{-1.0em}
  \label{fig:LosListDelete}
\end{figure}

\begin{figure}[!htp]
  \centering
  \vspace{-2.0em}
  \includegraphics[width=1\textwidth]{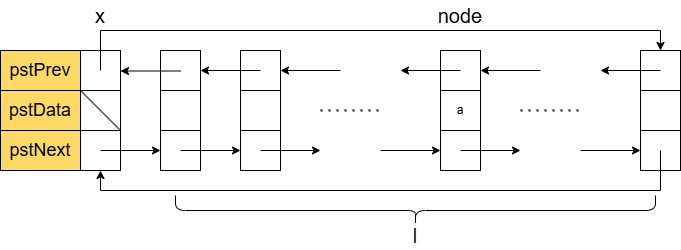}
  \vspace{-1.0em}
  \captionsetup{justification=raggedright, singlelinecheck=false}
  \caption{The definition of \lstinline{LOS_DL_LIST} in LiteOS. The node \lstinline{x} serves as a sentinel node, and its \lstinline{pstData} field remains unused.}
  \vspace{-2.0em}
  \label{fig:LOS_DL_LIST}
\end{figure}

For example, Figure~\ref{fig:LosListDelete} demonstrates the multiple specification case of the \lstinline{LOS_ListTailInsert} function from LiteOS~\cite{liteos_github}, a lightweight operating system based on a real-time kernel. To simplify the description, we have reduced the original polymorphic doubly-linked list to a circular doubly-linked list where each node stores an integer. Figure~\ref{fig:LOS_DL_LIST} illustrates the simplified definition of \lstinline{LOS_DL_LIST}, which consists of only a previous pointer (\lstinline{pstPrev}), a next pointer (\lstinline{pstNext}), and a data pointer (\lstinline{pstData}). The predicate \lstinline{store_dll(x)} represents a circular doubly-linked list storage structure, where \lstinline{x} denotes the sentinel node. In this context, our focus is solely on the overall shape property, while the actual data list is disregarded.


In this case, the \lstinline{LOS_ListAdd} function is equipped with two specifications. The \lstinline{primitive_spec} is used for the verification of \lstinline{LOS_ListAdd} itself as well as the \lstinline{LOS_ListTailInsert} function. Meanwhile, since \lstinline{LOS_ListAdd} can also function as a routine for inserting a node at the head, an additional \lstinline{head_insert_spec} is provided to describe this high-level behavior. QCP then requires the user to prove that \lstinline{head_insert_spec} can be derived from \lstinline{primitive_spec}, thereby eliminating the need for a separate, redundant verification of \lstinline{LOS_ListAdd} against \lstinline{head_insert_spec}.

\subsection{C function verification}\label{sec::cfunctionv}

As shown in previous examples in Figure~\ref{fig:sll_free} and~\ref{fig:LosListDelete}, users can use assertion annotations to verify a C function against its primitive specification. Given an input annotated C function, QCP performs symbolic execution to compute the strongest postcondition and generates corresponding verification conditions (VCs). These VCs are then processed by an entailment solver, which automatically discharges some verification conditions. The entailment solver integrates a rule-based abductive reasoning system (similar to the one used in VeriFast) and a lightweight SMT solver. Any remaining VCs are exported to Rocq for manual proof construction. This design combines the strengths of existing verification tools: it retains Rocq's expressiveness and proving power through manual proof engineering, while enhancing automation and efficiency through the entailment solver.

We now proceed to demonstrate the QCP verification process using the function in Figure~\ref{fig:sll_free}. Throughout the verification, it must be guaranteed that \lstinline{x->next} remains properly accessible for the store operation at line 14, which requires solving the following entailment:

\begin{lstlisting}
x != NULL && listrep(x) |-- exists v R, x->next == v && R    
\end{lstlisting}

\noindent At line 19, the user introduces a new assertion annotation to unfold the \lstinline{listrep(x)} predicate, ensuring that the symbolic execution of the read operation on \lstinline{x->next} at line 19 can proceed correctly. For this user-introduced annotation, QCP generates the following verification condition:

\begin{lstlisting}
    x != NULL && listrep(x) 
|-- exists v n, x -> next == n && x -> data == v && listrep(n)
\end{lstlisting}

This verification condition can be automatically discharged by our entailment solver. As symbolic execution progresses, the verification process generates multiple VCs, comprising both invariant validity checks and function call verifications. These VCs are systematically organized into four distinct output files:
\begin{itemize}[labelindent=1em]
    \item \lstinline{proof_goal.v}: Contains statements of all generated VCs, including the derivation of function specifications and VCs generated during the symbolic execution of function bodies.
    \item \lstinline{proof_auto.v}: Contains proof of all automatically verified VCs
    \item \lstinline{proof_manual.v}: Contains proof goals of VCs requiring manual verification.
    \item \lstinline{proof_check.v}: Ensuring all VCs present in \lstinline{proof_goal.v} are properly accounted for in either \lstinline{proof_auto.v} or \lstinline{proof_manual.v}
\end{itemize}
The files \lstinline{proof_goal.v}, \lstinline{proof_auto.v}, and \lstinline{proof_check.v} are automatically generated and require no user modification. The file \lstinline{proof_manual.v} contains only the VCs to be proved, which require manual completion of the proofs. In practice, we achieve full automatic verification for this example in Figure~\ref{fig:sll_free}, which can refer to Appendix~\ref{sec:sll_free_example}.

\section{Integration of development and incremental verification}\label{sec::VSCode}

Notably, our tool supports incremental verification and simultaneous development of a program and its correctness proof—it operates not only on fully annotated code but also presents intermediate verification results and program states through its VS Code IDE integration and web interface \textsc{Qide}. The interested readers could access \url{https://ide.qua.codes} to test it. Although the two interfaces differ slightly in layout, they present identical verification information. The following explanation will be based on the VS Code IDE version.

\begin{figure}[!htp]
    \vspace{-2.0em}
    \centering
    \includegraphics[width=\textwidth]{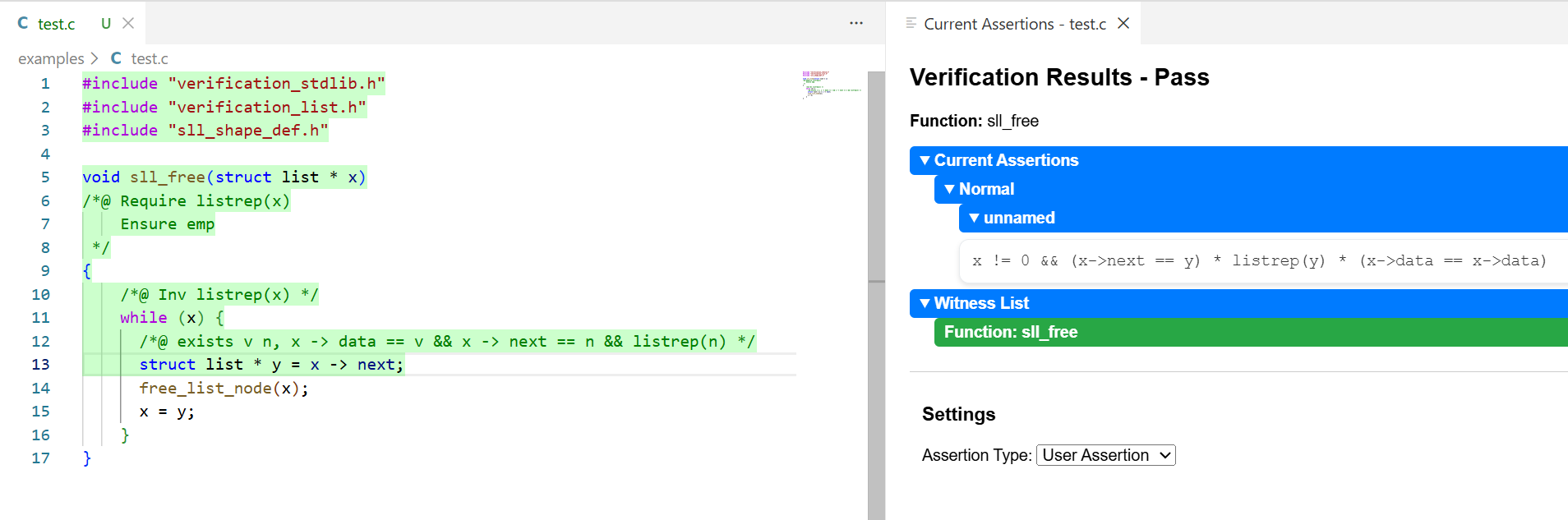}
    \vspace{-1.0em}
    \captionsetup{justification=raggedright, singlelinecheck=false}
    \caption{Live verification during code editing in VS Code.}
    \vspace{-2.0em}
    \label{fig:VSCode}
\end{figure}

As shown in the screenshot of Figure~\ref{fig:VSCode}, which corresponds to the \lstinline{sll_free} function in Figure~\ref{fig:sll_free}, the current symbolic state is displayed on the right-hand side. To enhance usability for developers with varying levels of expertise, we provide multiple assertion formats that allow users to customize how the current state is visualized and interpreted. Such detailed feedback helps users better understand program behavior and the verification process, thereby assisting them in completing annotations—or refining unfinished programs—more effectively. Green highlighting indicates code segments that have passed preliminary checks and are ready for symbolic execution to generate verification conditions. This demonstrates how our VS Code plugin allows users to seamlessly incorporate verification into their development workflow, offering real-time validation during code writing.

\begin{figure}[!htp]
    \vspace{-2.0em}
    \centering
    \includegraphics[width=\textwidth]{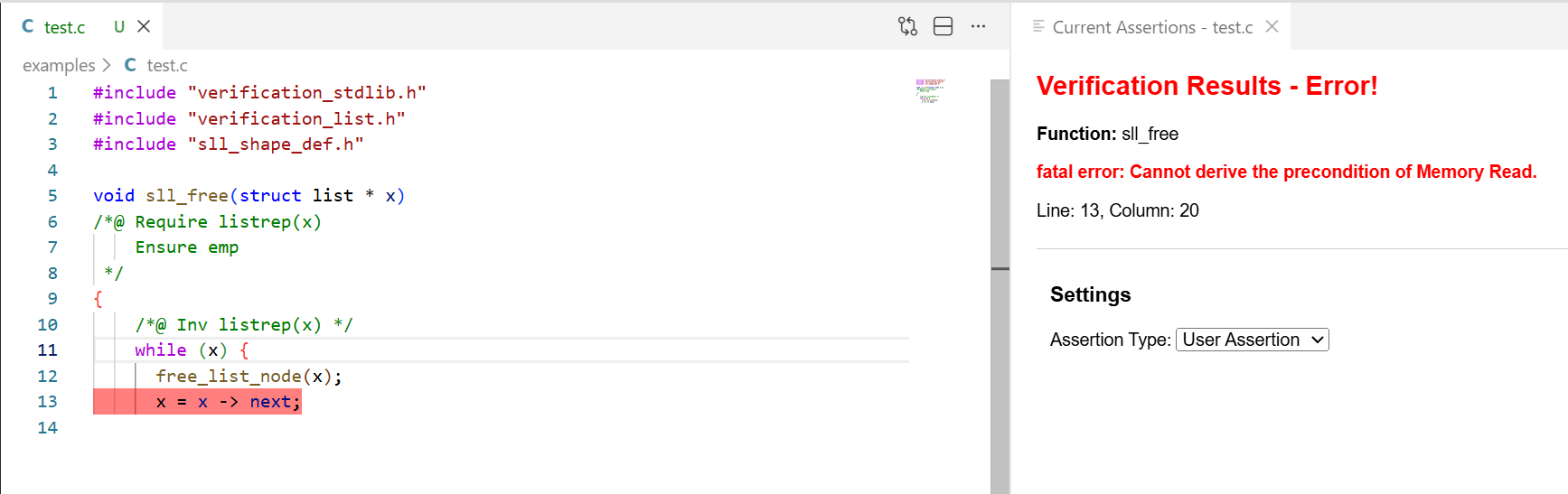}
    \vspace{-1.0em}
    \captionsetup{justification=raggedright, singlelinecheck=false}
    \caption{Failure case for live verification during code editing in VS Code.}
    \vspace{-2.0em}
    \label{fig:VSCode_fail}
\end{figure}

As for failure scenarios, Figure~\ref{fig:VSCode_fail} illustrates a case in which an incorrect implementation of the \lstinline{sll_free} function leads to a use-after-free error. In this situation, symbolic execution fails due to the inability to retain permission for \lstinline{x->next} after the memory has been freed. Within the VS Code interface, the affected code segment is highlighted in red on the left, while the right-hand panel displays the error message: “Cannot derive the precondition of Memory Read.” It is worth noting that the code in Figure~\ref{fig:VSCode_fail} is not yet complete. Even in such cases, QCP is able to provide intuitive and actionable error feedback, assisting users in diagnosing and correcting such faults.

For those cases that pass symbolic execution but cannot be fully solved automatically, we use yellow highlighting to distinguish them from failure cases. The corresponding code lines will be marked in yellow.

In summary, QCP features an integrated development interface that provides immediate visual feedback during the verification process. This significantly enhances the user's understanding of program behavior and proof status, thereby facilitating more efficient annotation writing and error debugging.
\section{Implementation} \label{sec::implementation}


\subsection{Symbolic execution}

QCP's frontend decomposes complete or incomplete C programs into what we term partial statements, which subsequently serve as the input to QCP’s symbolic executor. Figure~\ref{fig:state_fig} illustrates the decomposition of an incomplete function into its constituent partial statements, with different types of partial statements distinguished by color. These include: one program variable declaration \lstinline{(struct list *y)}; two singleton statements (specifically, the assignments \lstinline{y = x->next} and the function call \lstinline{free_list_node(x)}; two annotation statements;
one control-flow statement (\lstinline{break}); one conditional statement(\lstinline{if (x -> data == 0)});   one iteration statement (\lstinline{while}); two block begin and a block end. The complete grammar for partial statements is provided in Figure~\ref{fig:CPstatement}.

\begin{figure}[!htbp]
    \centering
    \vspace{-1.0em}
    \begin{outcodebox}
    \annotationbox{/*@ Inv listrep(x) */} \\
    \iterbox{while (x != NULL)} \blockbox{\{}
\begin{innercodebox}
\annotationbox{/*@ exists v n, x -> data == v \&\& x -> next == n \&\& listrep(n) */} \\
\declbox{struct list *y;} \singlebox{y = x->next;} \\
 \condbox{if (x -> data == 0)} \blockbox{\{} \ctrlbox{break;} \blockbox{\}} \\
 \singlebox{free\_list\_node(x);}
\end{innercodebox}
\end{outcodebox}
\vspace{1.0em}
\captionsetup{justification=raggedright, singlelinecheck=false}
    \caption{Example illustrating symbolic state during partial program execution, with different types of partial statements distinguished by color.}
    \vspace{-1.0em}
    \label{fig:state_fig}
\end{figure}

\begin{figure}[!htbp]
\begin{tabular}{llcll}
  Partial Statement & $ps$ & ::= & $s$ & \quad singleton statement\\
  && \myalt & $asrt$ & \quad annotation statement \\
  && \myalt & break \myalt continue & \quad control-flow statement \\
  && \myalt & return \myalt return $e$ & \quad return statement \\
  && \myalt & if ($e$) \myalt else & \quad conditional-if/else \\
  && \myalt & switch($e$) \myalt case $e$: \myalt default: & \quad switch-case-default \\ 
  && \myalt & while ($e$) \myalt do \myalt for($s$;$e$;$s$) & \quad iteration\\
  && \myalt & $\{$ \myalt $\}$ & \quad block begin/end  \\
  && \myalt & $ctype$ var\_name & \quad program variable declaration
\end{tabular}
\captionsetup{justification=raggedright, singlelinecheck=false}
\caption{Syntax of C partial statements. Singleton statements encompass assignments, increment/decrement operations, and function calls.}

\label{fig:CPstatement}
\end{figure}

The symbolic execution over partial statements follows conventional methodologies. During execution, QCP maintains a symbolic state stack. Figure~\ref{fig:state_fig} shows a partial program example: when execution reaches point \lstinline{free_list_node(x);}, the stack contains two nodes as shown by two rectangular blocks. Within the inner block, QCP maintains both a \texttt{Normal} assertion (when \lstinline{x -> data} does not store 0) and a \texttt{Break} assertion (when \lstinline{x -> data} stores 0). Similarly, \lstinline{continue} and \lstinline{return} statements generate \texttt{Continue} and \texttt{Return} assertions, respectively. These states are visualized in real-time within the \textsc{Qide} environment (see Figure~\ref{fig:VSCode}). Upon loop exit, the four types of inner assertions—\texttt{Normal}, \texttt{Break}, \texttt{Continue}, and \texttt{Return}—are merged with the outer assertion before symbolic execution continues with subsequent instructions. 

\subsection{Entailment solver}

To efficiently discharge verification conditions, QCP employs a custom entailment solver at its core. This solver is specifically designed to automate the reasoning process within our verification framework. As shown in Figure~\ref{fig:QCPframe}, the solver integrates an abductive reasoning system with a lightweight SMT solver.

The reasoning system is a rule-based separation logic engine responsible for eliminating spatial constraints in separation logic formulas. It supports user-defined predicates and corresponding elimination rules while simultaneously generating soundness proofs for each deduction step. QCP's implementation builds upon the Stellis framework~\cite{wang2025stellisstrategylanguagepurifying}, and the details are provided in Appendix~\ref{sec:Stellis}.

Our SMT solver implements decision procedures for two fundamental theories: \textbf{Linear Integer Arithmetic (LIA)}~\cite{DANTZIG1973288} and \textbf{Uninterpreted Functions (UF)}~\cite{10.1007/978-3-540-32033-3_33}. The solver architecture follows the Nelson-Oppen framework~\cite{10.5555/1391237} for theory combination, ensuring soundness when processing formulas containing mixed LIA and UF constraints. Currently, the solver does not support quantified formulas; such cases require manual user intervention.

\subsection{Rocq library}

QCP formalizes a C memory model and develops a separation logic library upon it. This library includes a C-specific assertion language, support for relational reasoning, and a suite of specialized proof tactics. The library employs C-style notations that closely resemble front-end C assertions, while being expressed in standard separation logic form. For example, the verification condition discussed in Section~\ref{sec::cfunctionv} is presented as follows:

\begin{lstlisting}
 forall (x: Z), 
  [| (x <> 0) |] &&  (listrep x)
|-- EX (v: Z)  (n: Z),
  [| (x <> 0) |] && ((&((x)  # "list" -> "next")) # Ptr mapsto n)
  **  (listrep n) ** ((&((x)  # "list" -> "data")) # Int mapsto v)
\end{lstlisting}

To efficiently manage the proof burden associated with separation logic, QCP provides a suite of specialized proof tactics. These tactics handle the tedious aspects of reasoning about spatial assertions, such as:
\begin{itemize}
   \item \lstinline{Intros/Exists}: Introduces or instantiates existentially quantified variables, and improves the organization of canonical preconditions.
    \item \lstinline{entailer!}: Automatically cancels common spatial predicates on both sides of the entailment by leveraging the associativity and commutativity of the separating conjunction.
    \item \lstinline{sep_apply}: Performs separation logic transformations by applying lemmas; it automatically collects the left spatial predicates of a lemma against those in the entailment’s left side, replacing them with the lemma’s right side.
\end{itemize}

Furthermore, QCP incorporates relational verification capabilities through Shushu Wu's new discovery that can encode forall-exists relational Hoare logic into standard Hoare logic~\cite{10.1145/3763138}. This design allows users to write relational specifications for conducting refinement proofs and algorithm consistency verification within the same framework. The system smoothly supports standard Hoare logic, relational Hoare logic, and even hybrid reasoning styles, providing flexibility for diverse verification requirements. Several verification examples demonstrating these capabilities will be presented in Appendix~\ref{sec:relationalexample}.

\section{Evaluation}\label{sec::eval}

\subsection{Performance of QCP}


We evaluated QCP across 155 functions spanning five domains: arithmetic (Arith), typical data structures (Typical DS, covering sll/dll/trees/arrays), typical algorithm implementations (Typical alg, covering merge sort, Knuth-Morris-Pratt algorithm, etc.), LiteOS microkernel components (mainly for doubly-linked lists), QCP implementation components (QCP Impl, covering Fourier–Motzkin elimination algorithm~\cite{DANTZIG1973288}, Typeinfer algorithm, etc.). 

Our evaluation focuses on QCP's ability to verify imperative implementations that exhibit typical programming patterns rather than verification–oriented implementations. The supported C features (see Section~\ref{eval::C_feature}) are representative of these patterns and are exercised across all benchmark examples. The functions in QCP implementation components represent a codebase that evolved organically over a year of QCP's development, providing a case study of verifying non-verification-oriented software; details are provided in Appendix~\ref{real_alg}.

\begin{table*}[ht]
\vspace{-2.0em}
  \caption{Evaluation results for QCP, including annotation lines, the number of verification conditions (Auto VCs and Manual VCs), symbolic-execution time per function (SE Time), and Rocq checking time per function (Rocq Time).}
  
\begin{center}
\scalebox{0.9}{
    \begin{tabular}{|c|c|c|c|c|c|c|c|}
\hline
\multirow{2}{*}{Type} & \multirow{2}{*}{Functions} & Annotations & Total Code & \multirow{2}{*}{Auto VCs} & \multirow{2}{*}{Manual VCs} & SE Time & Rocq Time \\ 
    &  & (Lines) & (Lines) &  &  & (s) & (s)
                          \\ \hline
      Arithm   &   12   &  98  & 159 & 84 & 39 & 0.07 & 1.43  \\ \hline
      Typical DS  & 67 &  954  &  1135  & 716  & 192 & 0.40 & 2.38 \\ \hline
      Typical alg   & 19  &  574  & 339 & 214 & 101 &  0.06 & 3.23 \\ \hline
      QCP Impl & 36 & 1757 & 755 & 859 & 287 & 2.01 & 16.48 \\ \hline  
      LiteOS DLL   & 21 & 493 & 584 & 91 & 60 & 0.02 & 11.00 \\ \hline
Total & 155 & 3876 & 2972 & 1964 & 679 & 0.65 & 6.85 \\ 
\hline 
\end{tabular}
}
\end{center}
\vspace{-1.0em}
  \label{table:evaluation1}

\end{table*}

Table~\ref{table:evaluation1} reveals two key trends: (1) The code-to-annotation ratio ranges from 1:0.6 for arithmetic programs (159 code vs. 98 annotations) to 1:2.3 for QCP implementation components (QCP Impl), reflecting the intrinsic specification needs of imperative programming; (2) QCP automates 74.3\% of verification conditions (1964 auto vs. 679 manual VCs), achieving high automation in arithmetic checks while reserving expert effort in LiteOS (60.3\% automation). 

Several examples in our evaluation are verified fully automatically, and the results for these cases are presented in Table~\ref{table:evaluation2}. In these cases, QCP achieves automated verification with an average code-to-annotation ratio of 1:0.5.

\begin{table*}[ht]

\vspace{-1.0em}
  \caption{Automatically verified results for QCP}

\begin{center}
    \begin{tabular}{|c|c|c|c|c|c|c|}
\hline
\multirow{2}{*}{File} & \multirow{2}{*}{Type} & \multirow{2}{*}{Functions} & Annotations & Total Code & \multirow{2}{*}{Auto VCs} & SE Time \\ 
    & & & (Lines) & (Lines) &  & (s)
                          \\ \hline
      add  &  Arith & 5   &  27  & 27 & 18 & < 0.01  \\ \hline
      max3  & Arith & 1 &  9  & 14  & 4 & 0.01 \\ \hline
      sll\_auto & Typical DS  & 6 & 45  &  110  & 48 &  0.02  \\ \hline
      dll\_auto & Typical DS  & 9 & 66 & 187 & 81 & 0.16 \\ \hline
avl\_insert & Typical DS & 8 & 99 & 116 & 250 & 2.84 \\ \hline 
array\_auto & Typical DS & 9 & 68 & 85 & 81 & 0.03 \\ 
\hline 
\end{tabular}
\end{center}
\vspace{-1.0em}
  \label{table:evaluation2}
\end{table*}

\subsection{Comparison with separation-logic-based annotation verifiers}

Real-world programs often involve complex memory structures that necessitate the use of separation logic. However, some annotation-based verification tools, such as Frama-C~\cite{Kirchner2015FramaC}, do not support separation logic. For this reason, our evaluation is limited to a comparison with other separation logic-based verifiers. Table~\ref {table:evaluation_VF} and Table~\ref {table:evaluation_Hip} present a comparative evaluation of QCP against VeriFast and Hip/Sleek on a suite of fully automatic verification cases, measuring the number of required annotations and verification time. The data, drawn from their respective benchmarks, show that QCP achieves verification times comparable to VeriFast and Hip/Sleek in the fully automatic mode. Our evaluation can be found at \url{https://github.com/QinxiangCao/QualifiedCProgramming/releases/tag/v1.0}.

\begin{table*}[htbp]
\vspace{-2.0em}
\centering
\caption{Comparison of verification effort and verification checking time (Annotation Lines, Proof Lines, and Time in s) between QCP and VeriFast.}
\vspace{1.0em}
\label{table:evaluation_VF}
\begin{tabular}{|c|c|c|cc|cc|}
\hline
\multicolumn{3}{|c|}{} & \multicolumn{2}{c|}{QCP}  & \multicolumn{2}{c|}{VeriFast}  \\ \hline
Program & Functions & LoC & Anno. & Time(s) & Anno. & Time(s) \\ \hline
arraylist & 7 & 74 & 33 & 0.37 & 31 & 0.10 \\ \hline
strlib & 4 & 38 & 31 & 0.14 & 17 & 0.10 \\ \hline
sll\_stack & 9 & 77 & 40 & 0.09 & 46 & 0.17  \\ \hline
\end{tabular}

\end{table*}

\begin{table*}[htbp]
\vspace{-1.0em}
\centering
\caption{Comparison of verification checking time between QCP and Hip/Sleek.}
\vspace{1.0em}
\label{table:evaluation_Hip}
\begin{tabular}{|c|c|c|c|c|c|}
\hline
Program               & Functions & LoC & Anno. & Hip/Sleek Time(s) & QCP Time(s) \\ \hline
Singly linked list    & 16        & 139 & 32    & 0.93              & 0.05  \\ \hline
Doubly linked list    & 15        & 181 & 34    & 0.90              & 0.14  \\ \hline
Sort algorithms       & 10        & 110 & 20    & 1.52              & 0.05  \\ \hline
\end{tabular}
\vspace{-1.0em}
\end{table*}

\subsection{Comparison with interactive verifiers}

To the best of our knowledge, limited prior research has integrated annotation-based and interactive verification methodologies such as QCP. Among existing approaches, VST-A is the most relevant. We evaluate the performance of QCP against both VST-A and VST using the VST-A benchmark (see Table~\ref{table:evaluation3}). The benchmark data are sourced directly from the original publication~\cite{10.1145/3632911}. We recorded the number of annotations, proof codes, and verification checking time for each tool, including both the time required to generate Rocq code and the time to execute the Rocq proof.  

In terms of proof effort, QCP also significantly reduces the required proof lines by 52.8\% compared to VST-A (457 vs. 969) and 62.3\% compared to VST (457 vs. 1212), which is attributed to its use of an entailment solver that automates significant portions of the proof process. As observed in Table~\ref{table:evaluation1}, the symbolic execution time in QCP is notably lower than the corresponding Rocq verification time. Therefore, reducing proof length directly contributes to shorter Rocq execution time. For example, in the SLL category, QCP achieves a 95.4\% reduction in execution time compared to VST-A (16.11s vs. 349.75s) and an 83.3\% reduction compared to VST (16.11s vs. 96.39s).

It should be noted that QCP requires a larger number of annotations than VST-A. This is because QCP utilizes annotations to drive assertion transformations—effectively shifting the separation logic transformations that would traditionally be done within proofs into the annotation phase. Despite this increase in annotations, the combined total of annotation lines and proof lines in QCP remains lower than that of VST-A. This result demonstrates that QCP not only provides a more intuitive proving experience by allowing users to perform assertion transformations through manual annotations, but also achieves an overall reduction in the total verification effort.

\begin{table*}[htbp]
\vspace{-2.0em}
\centering
\caption{Comparison of verification effort and verification checking time (Annotation Lines, Proof Lines, and Time in s) among QCP, VST-A, and VST.}
\vspace{1.0em}
\label{table:evaluation3}
\begin{tabular}{|c|c|c|ccc|ccc|cc|}
\hline
\multicolumn{3}{|c|}{} & \multicolumn{3}{c|}{QCP} & \multicolumn{3}{c|}{VST-A} & \multicolumn{2}{c|}{VST} \\ \hline
Program & Function & Codes & Anno. & Proofs & Time(s) & Anno. & Proofs & Time(s) & Proofs & Time(s) \\ \hline
Basics & 8 & 78 & 59 & 8 & 0.65 & 61 & 84 & 84.36 & 156 & 15.90 \\ \hline
SLL & 18 & 350 & 209 & 457 & 16.11 & 140 & 969 & 349.75 & 1212 & 96.39 \\ \hline
DLL & 4 & 95 & 36 & 70 & 4.33 & 32 & 171 & 155.25 & 213 & 29.91 \\ \hline
BST & 4 & 115 & 93 & 169 & 13.62 & 67 & 202 & 190.35 & 364 & 33.83 \\ \hline
\end{tabular}
\vspace{-3.0em}
\end{table*}

\subsection{Supported C features} \label{eval::C_feature}

QCP provides comprehensive support for standard C types, encompassing integers, pointers, arrays, enumerated types, \lstinline{struct}, \lstinline{union}, and \lstinline{typedef}. The unsupported features consist of floating-point types, function pointers, and bit-fields.

Regarding expression handling, QCP accommodates most C expressions, including:
postfix operators (like array subscripting and structure member selection), arithmetic operators (both value and pointer arithmetic), bitwise and logical operators, conditional operator, and assignment operators.
The system strictly enforces C-standard implicit type conversions and implements short-circuit evaluation for all expressions. Notable unsupported features include string literals, the comma operator, and compound operators (e.g., $(\text{int}\ [\ ])\{2, 4\}$).

For control flow, QCP supports all standard C control flow statements including conditional branches (\lstinline{if-else}), iteration constructs (\lstinline{while}, \lstinline{for}, and \lstinline{do-while}), and selection statements (\lstinline{switch-case}). The only non-supported control flow feature is the \lstinline{goto} statement.
\section{Related work}\label{sec::related}

\paragraph{C module verification: multiple specifications in other tools}

Multiple specification verification is valuable in real-world program verification, and many verifiers have recognized this need and introduced relevant support. For instance, Frama-C~\cite{Kirchner2015FramaC} and VeriFast~\cite{VeriFast} allow multiple specifications to be verified simultaneously within a single function, without requiring these specifications to be derivationally related—an approach that is particularly suitable for independent multiple specifications. Although similar effects can be encoded through specification derivation (Appendix~\ref{appendix:multispec} will show an example), QCP will also incorporate such functionality in future development. Similar to specification derivation, VSU~\cite{10.1007/978-3-030-72019-3_5} enables specification multiplicity between modules via subsumption, yet each module internally maintains only one specification. In this regard, Hip/Sleek~\cite{10.5555/1763048.1763074} goes further: it supports simultaneous symbolic execution for multiple specifications and specification derivation~\cite{4404760}. But it requires all derivations to be automatically verified—a constraint that can become a bottleneck in complex correctness proofs. In contrast, QCP offers more flexible support for specification derivation by allowing manual proof of the derivation verification conditions.

\paragraph{C function verification: fully automated systems}

Tools like CBMC~\cite{kroening2023cbmccboundedmodel} employ bounded model checking to verify memory safety and undefined behavior, yet their reliance on solvers like Z3 leads to non-negligible false positives. Similarly, abstract interpretation tools such as SPARTA~\cite{SPARTA} and MemCAD~\cite{10.1007/978-3-319-57288-8_15} utilize various abstract domains to verify numerical and basic memory properties efficiently. While these fully automated tools excel at proving a wide range of simple properties, they generally fall short in establishing the full functional correctness of real-world programs.

\paragraph{C function verification: annotation-based system}

VeriFast~\cite{VeriFast} and Hip/Sleek~\cite{10.5555/1763048.1763074} are separation logic-based annotation verifiers, and CN~\cite{10.1145/3571194} is a separation logic-based refinement type annotation verifier. All three tools incorporate specialized annotations designed to assist in the symbolic execution process. They rely on SMT solvers or type checkers to discharge verification conditions, and for moderately complex cases, VeriFast and Hip/Sleek further allow users to supply lemma functions to aid the SMT solver. While this approach can support automated proofs of functional correctness to a certain extent, it remains insufficient for more complex verification scenarios. As demonstrated in Tables~\ref{table:evaluation_VF} and~\ref{table:evaluation_Hip}, QCP achieves verification automation comparable to that of VeriFast and Hip/Sleek. Moreover, by enabling manual proofs for verification conditions that cannot be resolved automatically, QCP can address more complex verification tasks.

As for Frama-C~\cite{Kirchner2015FramaC}, it is built upon first-order logic and does not support separation logic. While Frama-C excels at scalable, automated analysis of runtime errors and functional properties within its supported logic fragment, it is not convenient for verifying complex data structures and algorithms.




\paragraph{C function verification: interactive systems}

VST-A~\cite{10.1145/3632911} decomposes the entire annotated program into multiple straight-line programs in Rocq, requiring users to perform manual symbolic execution proofs using tactics. The entire system is built on CompCert, with corresponding tactics and decomposition operations formally verified for soundness in Rocq. In contrast, QCP only requires users to prove specific VCs while leveraging entailment solvers to automatically resolve others. This approach significantly reduces the user's workload and enhances verification efficiency.

RefinedC~\cite{Sammler2021RefinedC} translates annotated C programs directly into Rocq. It employs typing rules to drive execution and generate typing derivation VCs, which are mostly discharged through a library of Rocq tactics—though unresolved cases still require manual intervention. However, its lack of SMT-based automation limits proof power, while its full implementation within Rocq results in lower verification efficiency and higher usability overhead compared to QCP.


\section{Conclusion}\label{sec::conclu}

In this work, we presented QCP, a verification tool designed to efficiently verify complex program properties. By leveraging separation logic, QCP provides a flexible and intuitive framework for symbolic execution and verification condition generation. QCP integrates both entailment solvers and Rocq, balancing automation with the ability to handle intricate proofs manually when necessary. Our evaluation demonstrates QCP's effectiveness in verifying programs with common data structures and highlights its ability to reduce manual effort.

Looking ahead, we aim to extend QCP to support advanced language features such as function pointers and goto statements. We also plan to enhance its usability for developers—for instance, by enabling integrated verification of multiple specifications. QCP represents a significant step forward in making program verification more accessible and practical for real-world software development. 
%
%
%
\bibliographystyle{splncs04}
\bibliography{fullbib}
%




\appendix
\renewcommand{\theHsection}{A\arabic{section}}
\section{Overview of Stellis} \label{sec:Stellis}

Stellis is a domain-specific language (DSL) designed to streamline the automation of separation logic entailment proofs. By enabling users to specify verification strategies, Stellis systematically reduces complex separation logic formulas (combining spatial and pure constraints) into pure logical forms that can be directly processed by constraint solvers. This transformation eliminates manual reasoning about heap-allocated structures while preserving soundness.

To ensure correctness, Stellis integrates a novel algorithm based on the Ramify Rule. For each user-defined strategy, the framework automatically generates a corresponding soundness lemma. The validity of the strategy is thereby reduced to proving this lemma, decoupling high-level strategy design from low-level proof obligations. This approach not only guarantees formal soundness but also empowers users to extend verification capabilities without requiring deep expertise in separation logic metatheory.

\begin{figure}[t]
\begin{tabular}{llll}
$\text{Priority}$ & $r$ & $::=$ & $n$ \\
$\text{Pattern term}$& $\hat{t}$ & $::=$ & $n\mid \mathord{?}x \mid x \mid
\field{\hat{t}}{field} \mid f(\hat{t}_1, \hat{t}_2, ...)$ \\
$\text{Pattern pure formula}$ & $\hat{p}$ & $::=$ & $\hat{t}_1 == \hat{t}_2 \mid
{\sim}\hat{p} \mid \hat{p}_1 \oplus \hat{p}_2 \mid P(\hat{t}_1, \hat{t}_2, ...)$ \\
$\text{Pattern spatial formula}$ & $\hat{s}$ & $::=$ & $\textbf{emp} \mid
\store{\hat{t}_1}{\hat{t}_2} \mid A(\hat{t}_1, \hat{t}_2, ...)$ \\
$\text{Pattern formula}$&  $\hat{f}$ & $::=$ & $\hat{p} \mid \hat{s}$ \\
$\text{Left pattern}$& $q_l$ & $::=$ & $\hat{f}\text{ at } n$\\
$\text{Right pattern}$& $q_r$ & $::=$ & $\text{exists }\, x, q_r\mid \hat{f}\text{ at }
n$\\
$\text{Pattern}$ & $q$ & $::=$ & $\text{left: }q_l\mid \text{right: }q_r$ \\
$\text{Check}$ & $c$ & $::=$ & $\textbf{left\_absent}(p) \mid \textbf{right\_absent}(p) \mid \textbf{infer}(p)$\\
$\text{Operation}$ & $o$ & $::=$ & $\textbf{left\_add}(f) \mid
\textbf{right\_add}(f)$\\
&&& $\mid \textbf{left\_erase}(n) \mid \textbf{right\_erase}(n)$ \\
&&& $\mid \textbf{forall\_add}(x) \mid \textbf{right\_exist\_add}(x)$ \\
$\text{Action}$ & $a$ & $::=$ & $\vec{o} \mid \textbf{instantiate}(x\to t)$ \\
$\text{Strategy}$ & $S$ & $::=$ & $\text{priority: }r$\\
&&&$\vec{q}$\\
&&&$\text{check: }\vec{c}$\\
&&&$\text{action: }a$\\
$\text{Program}$ & $Prog$ & $::=$ & $\vec{S}$
\end{tabular}
\vspace{-1.0em}
\captionsetup{justification=raggedright, singlelinecheck=false}
    \caption{Syntax of Stellis}
    \vspace{-2.0em}
    \label{fig:strategy-syntax}
\end{figure}

Figure~\ref{fig:strategy-syntax} presents the formal syntax of Stellis. The figure and its accompanying description are from the original Stellis paper~\cite{wang2025stellisstrategylanguagepurifying}. An Stellis program $Prog$ consists of a sequence of strategies $\vec{S}$. A strategy $S$ has the following four elements:
\begin{enumerate}
    \item A priority $r$, specified using the \lstinline|priority| label.
    \item A sequence of patterns $\vec{q}$.
    \item A sequence of checks $\vec{c}$, denoted by the \lstinline|check| label.
    \item An action $a$, indicated by the \lstinline|action| label.
\end{enumerate}

The pattern part is used to identify specific formulas on both sides of the entailment. For right patterns, a pattern variable $x$ may be constrained to bind an existential variable in the entailment via the syntax ``$\text{exists } x, q_r$''. The pattern formula $\hat{f}$ follows the same syntactic structure as the formula $f$ in the entailment, except that $\hat{f}$ may contain $\mathord{?}x$, which introduce new pattern variables to bind to terms $t$ in the entailment.

The check part ensures the entailment satisfies specific constraints. For example, $\textbf{left\_absent}(p)$ confirms the absence of a pure formula $p$ in the antecedent, while $\textbf{infer}(p)$ invokes an SMT solver to determine whether a pure fact $p$ can be inferred from the antecedent.

The action part consists of two types: a sequence of operations $\vec{o}$ that manipulates the entailment by adding, removing, or introducing fresh variables, and $\textbf{instantiate}(x \to t)$, which instantiates an existential variable $x$ with a term $t$. 

We will present specific strategy examples in Appendix~\ref{sec:sll_free_example}.
\section{Examples: Full verification process for \lstinline{sll_free} in QCP} \label{sec:sll_free_example}

In the actual verification setup, as shown in Figure~\ref{fig:sll_free_real}, the process involves three header files (.h) and one implementation file (.c). The \lstinline{free_node_list} function is imported via \lstinline{sll_shape_def.h}. Upon completion, four output files are generated: \lstinline{sll_free_goal.v}, \lstinline{sll_free_proof_auto.v}, \lstinline{sll_free_proof_manual.v}, and \lstinline{sll_free_goal_check.v}. It should be noted that \lstinline{sll_free_proof_manual.v} remains an empty file in this process, as it can be automatically verified.

\begin{figure}[htp]
  \vspace{-2.0em}
  \centering
\begin{minted}{c}
#include "verification_stdlib.h"
#include "verification_list.h"
#include "sll_shape_def.h"

void sll_free(struct list * x)
/*@ Require listrep(x)
    Ensure emp
 */
{
    /*@ Inv listrep(x) */
    while (x) {
      /*@ exists v n, x -> data == v && x -> next == n && listrep(n) */
      struct list * y = x -> next;
      free_list_node(x);
      x = y;
    }
}
\end{minted}
  \vspace{-1.0em}
  \captionsetup{justification=raggedright, singlelinecheck=false}
  \caption{The real verification file of \lstinline{sll_free}. }
  \vspace{-2.0em}
  \label{fig:sll_free_real}
\end{figure}

Regarding proof checking, we leverage Rocq's Module Type mechanism. In \lstinline{sll_free_goal.v}, we generate a Module Type named \lstinline{VC_Correct} that contains all verification goals (see Figure~\ref{fig:VC_Correct_goal}), including the correctness proof of the strategies and the proof of the verification conditions. Subsequently, in \lstinline{sll_free_goal_check.v}, QCP automatically create a module \lstinline{VC_Correctness} of type \lstinline{VC_Correct} (see Figure~\ref{fig:VC_Correct_check}). This module includes all contents from both \lstinline{sll_free_proof_auto.v} and \lstinline{sll_free_proof_manual.v}, thereby utilizing Rocq to ensure that every goal has a corresponding proof.

\begin{figure}[!htp]
    \centering
    \vspace{-2.0em}
    \begin{lstlisting}
Module Type VC_Correct.
    Include common_Strategy_Correct.
    Include sll_shape_Strategy_Correct.

    Axiom proof_of_sll_free_entail_wit_1 : sll_free_entail_wit_1.
    Axiom proof_of_sll_free_entail_wit_2 : sll_free_entail_wit_2.
    Axiom proof_of_sll_free_entail_wit_3 : sll_free_entail_wit_3.
    Axiom proof_of_sll_free_return_wit_1 : sll_free_return_wit_1.
    Axiom proof_of_sll_free_partial_solve_wit_1 : sll_free_partial_solve_wit_1.
End VC_Correct.
\end{lstlisting}
\vspace{-1.0em}
    \caption{\lstinline{VC_Correct} Module Type for \lstinline{sll_free} in \lstinline{sll_free_goal.v}.}
    \vspace{-2.0em}
    \label{fig:VC_Correct_goal}
\end{figure}

\begin{figure}[!htp]
    \centering
\begin{lstlisting}
Module VC_Correctness : VC_Correct.
  Include common_strategy_proof.
  Include sll_shape_strategy_proof.
  Include sll_free_proof_auto.
  Include sll_free_proof_manual.
End VC_Correctness.
\end{lstlisting}
\vspace{-1.0em}
\captionsetup{justification=raggedright, singlelinecheck=false}
    \caption{\lstinline{VC_Correctness} Module for \lstinline{sll_free} in \lstinline{sll_free_proof_check.v}.}
    \vspace{-2.0em}
    \label{fig:VC_Correct_check}
\end{figure}

Although this example can be fully automatically solved by our solver, here we present the Rocq proof for the verification condition discussed in Section~\ref{sec::cfunctionv}, which is \lstinline{sll_free_entail_wit_2}. QCP provides a \lstinline{pre_process} command for preprocessing verification conditions. After preprocessing, the proof becomes :

\begin{lstlisting}
x : Z
H : x <> 0
----------------------------
listrep x 
|-- EX v : Z, EX n : Z,
    [| x <> 0 |] && 
    &( x # "list" -> "next") # Ptr |-> n ** listrep n ** 
    &( x # "list" -> "data") # Int |-> v
\end{lstlisting}

Next, given that \lstinline{x <> 0}, which means \lstinline{listrep(x)} is non-empty, we can use \lstinline{sep_apply listrep_nonzero} to unfold \lstinline{listrep(x)} into its non-empty form. The proof then becomes:

\begin{lstlisting}
x :Z
H: x <> 0
----------------------------
EX y : Z, EX a : Z,
&( x # "list" -> "data") # Int |-> a ** 
&( x # "list" -> "next") # Ptr |-> y ** listrep y
|-- EX v: Z, EX n: Z,
    [|x <> 0|] &&
    &( x # "list" -> "next") # Ptr |-> n ** listrep n **
    &( x # "list" -> "data") # Int |-> v
\end{lstlisting}

Subsequently, by using \lstinline{Intros y a} and \lstinline{Exists a y}, we introduce the existential quantifier on the left-hand side and instantiate the one on the right-hand side. Finally, applying \lstinline{entailer!} eliminates the separation logic components and completes the proof. Total proof scripts are shown below :
\begin{lstlisting}
Lemma proof_of_sll_free_entail_wit_2 : sll_free_entail_wit_2.
Proof.
  pre_process.
  sep_apply listrep_nonzero.
  Intros y a. Exists a y.
  entailer!.
Qed.
\end{lstlisting}

At the end of this section, we present the Stellis strategy we introduced to solve this verification condition automatically. This strategy is applied only when the left-hand side of the verification condition contains \lstinline{listrep(p)} and \lstinline{p != NULL}, and the right-hand side involves \lstinline{p->data} or \lstinline{p->next}. It works by expanding the \lstinline{listrep(p)} on the left-hand side. Here, the notation \lstinline{?p} indicates that the matched expression is bound to \lstinline{p}, and this is not limited to variable scenarios. In the \lstinline{left} and \lstinline{right} patterns, we use at notation to simplify the matched expressions. This approach ensures that the \lstinline{listrep} predicate is expanded only when load/store is required, avoiding unnecessary operations on unrelated predicates, thereby simplifying the assertion.

\begin{lstlisting}
id: 8
priority: core(6)
left: listrep(?p : Z) at 0
       (p != NULL || NULL != p) at 1
right: (data_at(field_addr(p, list, data), I32, ?q : Z) || data_at(field_addr(p, list, next), PTR(struct list), ?q : Z)) at 2
action: left_erase(0);
        left_exist_add(x : Z);
        left_exist_add(y : Z);
        left_add(data_at(field_addr(p, list, data), I32, x));
        left_add(data_at(field_addr(p, list, next), PTR(struct list), y));
        left_add(listrep(y));
\end{lstlisting}

\section{Example: language design comparison between QCP and RefinedC} \label{sec:rcexamples}

\begin{figure}[!ht]
    \centering
    \newsavebox\picA
    \savebox\picA{\includegraphics[width=0.45\textwidth]{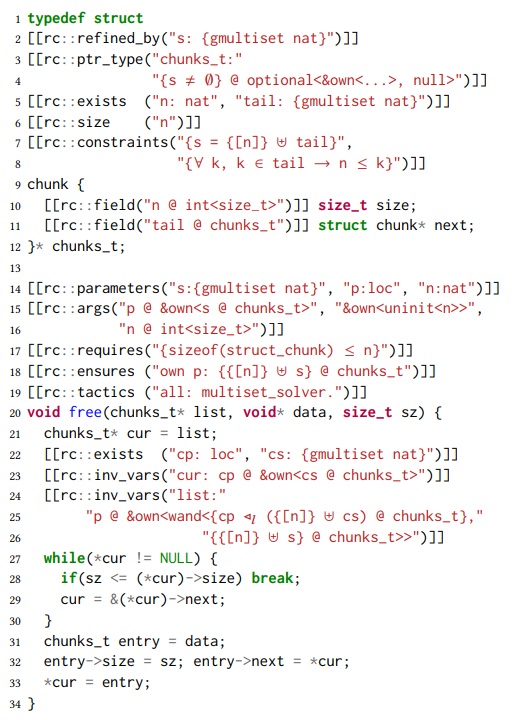}}
    
    \newsavebox\picB
    \savebox\picB{\includegraphics[width=0.45\textwidth]{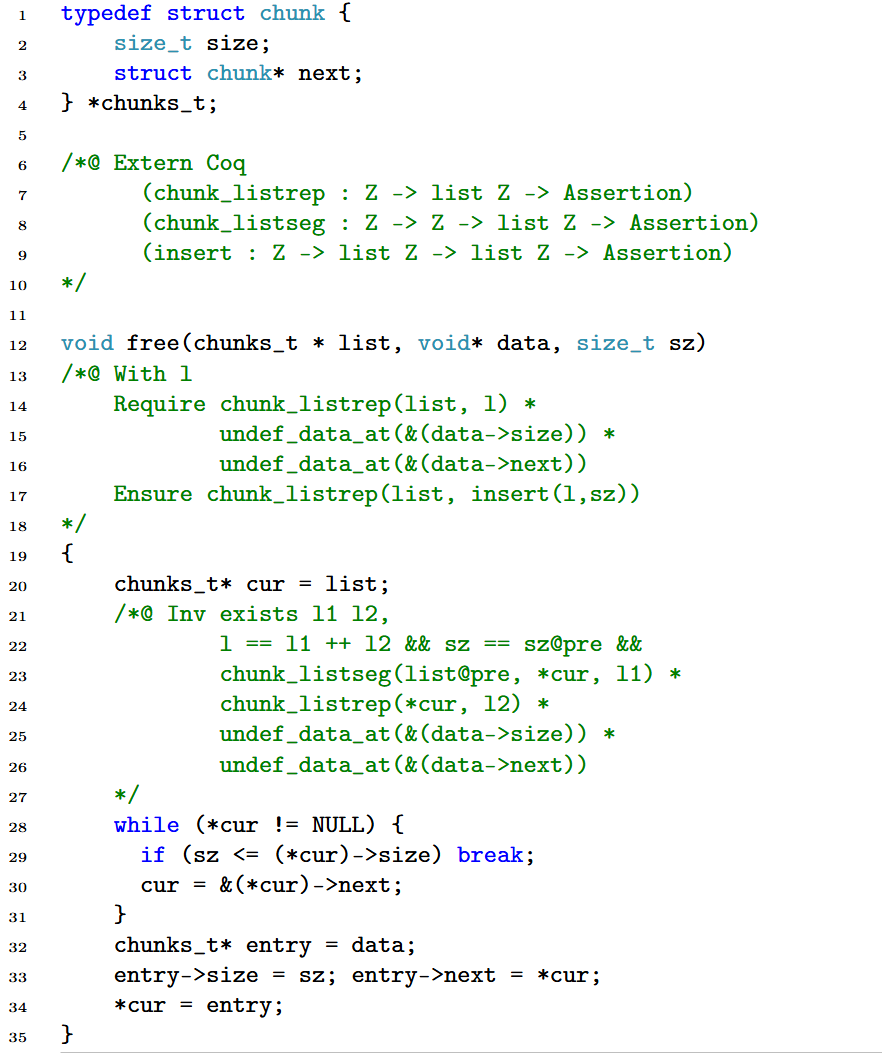}}
    
    \newlength\maxheight
    \setlength\maxheight{\maxof{\ht\picA}{\ht\picB}}
    \begin{minipage}[b]{0.45\textwidth}
        \centering
        \includegraphics[height=\maxheight, width=\linewidth, keepaspectratio]{Figures/RefinedC.png}
        \caption*{(a) RefinedC example}
    \end{minipage}
    \hfill
    \begin{minipage}[b]{0.5\textwidth}
        \centering
        \includegraphics[height=\maxheight, width=\linewidth, keepaspectratio]{Figures/free.png}
        \caption*{(b) Version in QCP}
    \end{minipage}
    \captionsetup{justification=raggedright, singlelinecheck=false}
    \caption{Language Design comparison between RefinedC and QCP.}
    \vspace{-1.0em}
    \label{fig:RefinedC}
\end{figure}

This example in Fig~\ref{fig:RefinedC} is adapted from page 5 of the RefinedC paper~\cite{Sammler2021RefinedC}. In comparison, QCP's annotation language is more intuitive and easier to read.

\section{Examples: Relational Proof in QCP} \label{sec:relationalexample}

We decompose the proof of functional correctness into two parts: the correctness of the algorithm itself and the consistency of its implementation. The former can be established directly in Rocq, while the latter can be naturally formulated and verified using relational Hoare logic—a proof system for reasoning about the equivalence or refinement between two programs. Within this framework, we verify a merge sort function by implementing an abstract program using monads in Rocq. This approach employs an imperative-style formulation, which more effectively captures the algorithmic behavior to be verified. Figure~\ref{fig:mergesortrec} illustrates this abstract program formulation for the merge sort function.

\begin{figure}[!htbp]
    \centering
    \vspace{-2.0em}
    \begin{lstlisting}
Definition merge_body:
  list Z * list Z * list Z ->
  MONAD (CntOrBrk  (list Z * list Z * list Z) (list Z)) :=
  fun '(l1, l2, l3) =>
    match l1, l2 with 
    | nil, _ => return (by_break (l3 ++ l2))
    | _, nil => return (by_break (l3 ++ l1))
    | x :: l1', y :: l2' =>
    choice
      (test' (x <= y);; return (by_continue (l1', l2, l3 ++ x :: nil)))
      (test' (y <= x);; return (by_continue (l1, l2', l3 ++ y :: nil)))
  end.

Definition merge_rel l l0 :=
    repeat_break merge_body (l, l0, nil).

Definition mergesortrec_f  (W :  (list Z) -> MONAD (list Z) ) 
                    : ((list Z) -> MONAD (list Z)) :=
fun x => '(p1, q1) <- split_rel x ;; 
                  match q1 with 
                  | nil => return p1
                  | _ :: _  =>  p2 <- W (p1) ;; 
                                q2 <- W (q1) ;; 
                                merge_rel p2 q2
                  end.

Definition mergesortrec := Lfix (mergesortrec_f).
\end{lstlisting}
\vspace{-1.0em}
    \captionsetup{justification=raggedright, singlelinecheck=false}
    \caption{Definition of \lstinline{mergesortrec}.}
    \vspace{-2.0em}
    \label{fig:mergesortrec}
\end{figure}

By applying the encoding introduced by Shushu Wu, which transforms $\forall - \exists$ relational Hoare logic into standard Hoare logic, QCP is employed to prove that the implemented code refines the algorithm described by the abstract program. We obtain the encoded specification shown in Figure~\ref{fig:relationalspec}. Here, \lstinline{safeExec(True, mergesortrec(l), X)} expresses that there exists an initial state satisfying \lstinline{True} such that any final state after executing \lstinline{mergesortrec(l)} satisfies \lstinline{X}. In essence, this specification can be interpreted as follows: given a list \lstinline{l} stored in a singly linked list at address \lstinline{x}, upon termination of the function, we have effectively executed the abstract program \lstinline{mergesortrec(l)} and returned \lstinline{l0}, which is then stored in the linked list at \lstinline{x}.

\begin{figure}
    \centering
    \begin{minted}{c}
struct list * merge_sort(struct list * x)
/*@ With l X
  Require safeExec(ATrue, mergesortrec(l), X) && sll(x, l)
  Ensure exists l0,
          safeExec(ATrue, return(l0), X) &&
          sll(__return, l0)
*/;
\end{minted}
\vspace{-1.0em}
\captionsetup{justification=raggedright, singlelinecheck=false}
    \caption{Relational specification of merge sort example.}
    \vspace{-1.0em}
    \label{fig:relationalspec}
\end{figure}

\section {Verification results for programs in QCP implementation components } \label{real_alg}.

\begin{table*}[!ht]
\centering
\vspace{-2.0em}
\caption{Verification results for QCP implementation components}
\scalebox{0.85}{
\begin{tabular}{|c|c|c|c|c|c|c|c|c|}
\hline
\multirow{2}{*}{File} & \multirow{2}{*}{Functions} & Anno. & Codes & \multirow{2}{*}{Auto VCs} & \multirow{2}{*}{Manual VCs} & Manual Proof & SE Time & Rocq Time \\ 
    & & (Lines) & (Lines) & & & (Lines) & (s) & (s)
                          \\ \hline
gmp\_add & 14 & 315 & 128 & 110 & 77 & 790 & 0.97 & 4.58 \\ \hline 
fme  & 9 & 253 & 165 & 142 & 38 & 522 & 3.59 & 9.78 \\ \hline 
typeinfer & 3 & 366 & 109 & 110 & 35 & 400 & 0.10 & 11.76 \\ 
\hline 
alpha\_equiv  & 1   &  105  & 33 & 35 & 22 & 228 & 1.01 & 45.58 \\ \hline
subst   & 2 &  147  & 61  & 42 & 32 & 194 & 0.12 & 14.36  \\ \hline
thm\_apply & 4 & 192  & 64  & 62 & 34 & 323 & 0.06 & 10.84 \\ \hline
cnf\_trans  & 3 & 379 & 195 & 358 & 49 & 1550 & 8.27 & 96.10 \\ \hline
\end{tabular}
}
\vspace{-2.0em}
\label{table:evaluation_real}
\end{table*}

Table~\ref{table:evaluation_real} presents the verification results for the QCP implementation components and Table~\ref{table:evaluation_goal} presents the verification properties for each file. In the following, we describe each algorithm individually and summarize the corresponding verification outcomes. We omit the detailed Rocq definitions and present only the specifications of the principal functions in each file.

\begin{table*}[!ht]
\centering
\caption{Verification Properties for QCP Implementation Components}
\begin{tabular}{|p{2cm}|p{10cm}|}
\hline
\textbf{File} & \textbf{Verification Property} \\ \hline
gmp\_add & The \lstinline{mpn_add} correctly performs high-precision addition calculations. \\ \hline 
fme & If \lstinline{lia_deduction} says a linear arithmetic entailment is correct, it is indeed correct. \\ \hline 
typeinfer & The \lstinline{atype_unify} always finds the best unification solution. \\ \hline 
alpha\_equiv & The \lstinline{alpha_equiv} correctly determines alpha-equivalence. \\ \hline
subst & The \lstinline{subst} correctly computes symbolic substitution. \\ \hline
thm\_apply & The \lstinline{thm_apply} correctly preforms theorem application. \\ \hline
cnf\_trans & The \lstinline{cnf_trans} correctly performs Tseitin transformation. \\ \hline
\end{tabular}
\label{table:evaluation_goal}
\end{table*}

\paragraph{\lstinline{gmp_add}:}

The file \lstinline{gmp_add}, taken from the minigmp library, implements high-precision addition. We have formally verified the functional correctness of this algorithm.

\begin{minted}{c}
unsigned int mpn_add (unsigned int *rp, unsigned int *ap, 
                      int an, unsigned int *bp, int bn)
/*@ With val_a val_b
 Require an >= bn && an > 0 && bn > 0 &&
         mpd_store_Z(UINT_MOD, ap, val_a, an) *
         mpd_store_Z(UINT_MOD, bp, val_b, bn) *
         UIntArray::undef_full(rp, an)
 Ensure exists val_r_out,
   (val_r_out + __return * Z::pow(UINT_MOD, an) == val_a + val_b) &&
   mpd_store_Z(UINT_MOD, ap, val_a, an) *
   mpd_store_Z(UINT_MOD, bp, val_b, bn) *
   mpd_store_Z(UINT_MOD, rp, val_r_out, an)
*/    
\end{minted}

The store predicate for mini-gmp is \lstinline{mpd_store_Z}. The specification states that if \lstinline{val_a} (a large integer) is stored at address \lstinline{ap} with length \lstinline{an}, \lstinline{val_b} (a large integer) is stored at address \lstinline{bp} with length \lstinline{bn}, and \lstinline{an} is greater than \lstinline{bn}, then upon function completion, the result of the addition will be stored starting at address \lstinline{rp}, and the carry-out will be returned as the function's return value.

\paragraph{\lstinline{fme}:}

The \lstinline{fme} module implements the Fourier–Motzkin elimination algorithm, a quantifier elimination procedure used to solve systems of linear inequalities by iteratively removing variables and projecting the solution space onto remaining dimensions. 

\begin{minted}{c}
int lia_deduction(struct InequList** pr, int n)
/*@ With p1 l1
    Require BP0 != 0 && pr != 0 &&
            n >= 1 && n <= INT_MAX - 1 && 
            BP0->upper == 0 &&
            BP0->lower == 0 &&
            BP0->remain == 0 &&
            LP_abs_in_int_range(n+1, l1) &&
            data_at(pr, p1) *
            InequList(p1, n + 1, l1)
    Ensure ((__return == -1 &&
            data_at(pr, p1) *
            undef_data_at(&(BP0->upper)) *
            undef_data_at(&(BP0->lower)) *
            undef_data_at(&(BP0->remain))) 
            ||
            (exists p2, __return == 1 && UNSAT(l1) && 
                data_at(pr, p2) *
                undef_data_at(&(BP0->upper)) *
                undef_data_at(&(BP0->lower)) *
                undef_data_at(&(BP0->remain)))
            ||
            (exists p2 l2, __return == 0 && 
                LP_implies(l1, l2) && InequList_Zeros(l2, 1, n + 1) &&
                data_at(pr, p2) *
                InequList(p2, n + 1, l2) *
                undef_data_at(&(BP0->upper)) *
                undef_data_at(&(BP0->lower)) *
                undef_data_at(&(BP0->remain))))
  */
\end{minted}

Acting as the entry point for Fourier-Motzkin Elimination, the \lstinline{lia_deduction} function's return value has specific meanings: a value of -1 signals a failure to complete the reduction; a value of 0 means the reduction finished but did not produce a contradiction (i.e., cannot prove UNSAT); and a value of 1 signifies a successful reduction to a contradictory system, conclusively proving the original problem is UNSAT. 

The store predicate for inequations is \lstinline{InequList}. The specification states that if a system of inequalities \lstinline{l1} is stored at address \lstinline{p1} with n variables, and if every coefficient in \lstinline{l1} falls within the range of the int type, then this function will attempt to reduce the original inequality system according to the Fourier-Motzkin Elimination (FME) algorithm and return the corresponding result.

\paragraph{\lstinline{typeinfer}:}

The \lstinline{typeinfer} file implements a type inference algorithm, in which \lstinline{atype_unify} performs type unification—a core subroutine that resolves equality constraints between type expressions by computing a most general unifier or detecting inconsistency.

\begin{minted}{c}
int atype_unify(struct atype *t1, struct atype *t2)
/*@ With (s_pre : sol) (tr1 : TypeTree) (tr2 : TypeTree)
    Require 
        store_compressed_solution(res, s_pre) *
        store_type(t1, tr1) *
        store_type(t2, tr2)
    Ensure
        (
            (
            exists s_post,
                __return == 0 &&
                sol_correct_iter(tr1, tr2, s_pre, s_post) &&
                store_compressed_solution(res, s_post) *
                store_type(t1, tr1) *
                store_type(t2, tr2)
            )
            ||
            (
            exists s_post,
                __return != 0 &&
                store_solution(res, s_post) *
                store_type(t1, tr1) *
                store_type(t2, tr2)
            )
        )
*/  
\end{minted}

Below are the Rocq definitions of \lstinline{store_compressed_solution} and \linebreak \lstinline{sol_correct_iter}. The \lstinline{store_compressed_solution} predicate indicates that a solution \lstinline{s'} is stored in memory, where \lstinline{s} is the compressed representation of \lstinline{s'} and \lstinline{s} is guaranteed to be finite and acyclic. The \lstinline{sol_correct_iter} predicate expresses that the solution after unification is equivalent to the previous solution.

\begin{lstlisting}
Definition sol_correct_iter (t1 t2: TypeTree) 
                            (s_pre s_post: sol): Prop :=
forall sf: sol, 
    sol_compressed sf ->
    sol_valid_eq t1 t2 sf /\ sol_refine s_pre sf <-> sol_refine s_post sf.

Definition store_compressed_solution (x : addr) (s : sol) : Assertion :=
EX (s' : sol), 
  [| sol_compress_to s' s |] && [| sol_finite s |] && 
  [| sol_no_loop s' |] &&
  store_solution x s'.
\end{lstlisting}

For the \lstinline{atype_unify} function, a return value of 0 indicates that unification has completed successfully, and the unified solution is stored in the global variable \lstinline{res}. Any non-zero return value indicates that unification has failed.

\paragraph{\lstinline{alpha_equiv}:}

The \lstinline{alpha_equiv} file implements an algorithm for determining $\alpha$-equivalence between two terms—a fundamental operation in syntactic equality modulo bound variable renaming. The implementation of \lstinline{term_alpha_eqn} is similar to \lstinline{alpha_equiv}, so we have omitted the Rocq definitions here.

\begin{minted}{c}
bool alpha_equiv(term *t1, term *t2)
/*@ With term1 term2
  Require store_term(t1, term1) *
          store_term(t2, term2)
  Ensure __return == term_alpha_eqn(term1, term2) && 
         t1 == t1@pre && t2 == t2@pre && 
         store_term(t1, term1) * store_term(t2, term2)
*/
\end{minted}

The predicate \lstinline{store_term(t1, term1)} indicates that the term \lstinline{term1} is stored at address \lstinline{t1}. The specification states that if addresses \lstinline{t1} and \lstinline{t2} store terms \lstinline{term1} and \lstinline{term2} respectively, then the function will return whether the two terms are alpha-equivalent.

\paragraph{\lstinline{subst}:}

The \lstinline{subst} file provides a substitution algorithm that systematically replaces free variables within a term while preserving binding structure and avoiding variable capture. The implementation of \lstinline{term_subst_t} is similar to \lstinline{subst_term}, so we have omitted the Rocq definitions here.

\begin{minted}{c}
term *subst_term(term *den, char *src, term *t)
/*@ With trm src_str den_term
  Require den != 0 && src != 0 && t != 0 &&
          store_term(t, trm) *
          store_string(src, src_str) *
          store_term(den, den_term)
  Ensure den == den@pre && src == src@pre &&
         store_term(__return, term_subst_t(den_term, src_str, trm)) *
         store_term(den, den_term) *
         store_string(src, src_str)
*/
\end{minted}

\paragraph{\lstinline{thm_apply}:}

The \lstinline{thm_apply} component implements theorem application, a core routine that matches the conclusion of a given theorem against a target goal and appropriately instantiates its premises to construct a derivation.

\begin{minted}{c}
solve_res* thm_apply(term* thm, var_sub_list* lis, term* goal) 
/*@ With t l g X
  Require thm != 0 &&
          safeExec(ATrue, thm_app_rel(t, l, g), X) &&
          store_term(thm, t) * store_term(goal, g) *
          sll_var_sub_list(lis, l)
  Ensure exists sr t,
          thm == thm@pre &&
          safeExec(ATrue, return(sr), X) &&
          store_term(thm, t) * store_term(goal, g) *
          sll_var_sub_list(lis, l) * 
          store_solve_res(__return, sr)
*/
\end{minted}

In this example, we apply the method outlined in Appendix~\ref{sec:relationalexample} to construct an abstract program, \lstinline{thm_app_rel}, and subsequently prove the consistency of the algorithmic implementation.

\begin{lstlisting}
Definition thm_app: 
  term * var_sub_list * term ->
  MONAD (solve_res) :=
  fun '(t, l, g) =>
  match thm_subst_allres t l with
  | None => ret (SRBool 0)
  | Some (_, thm_ins) =>
      if (term_alpha_eq thm_ins g) then ret (SRBool 1)
      else x <- (check_rel thm_ins g) ;; get_list x
  end.

Definition thm_app_rel (thm : term) (l : var_sub_list) (goal : term) :=
  thm_app (thm, l, goal).
\end{lstlisting}

\paragraph{\lstinline{cnf_trans}:}

The \lstinline{cnf_trans} file implements the Tseitin transformation algorithm, which converts arbitrary propositional formulas into conjunctive normal form by introducing auxiliary variables and clauses while preserving satisfiability. The implementation of \lstinline{prop2cnf_logic} is similar to \lstinline{prop2cnf}, so we have omitted the Rocq definitions here.

\begin{minted}{c}
int prop2cnf(SmtProp *p, PreData *data)
/*@ With prop clist pcnt ccnt
  Require prop_cnt_inf_SmtProp(prop) <= pcnt &&
      SmtProp_size(prop) <= 10000 &&
      Zlength(clist) <= 40000 - 4 * SmtProp_size(prop) &&
      pcnt <= 40000 - SmtProp_size(prop) &&
      store_SmtProp(p, prop) *
      store_predata(data, clist, pcnt, ccnt)
  Ensure exists clist' pcnt' ccnt' res,
     make_prop2cnf_ret(make_predata(clist', pcnt', ccnt'), res) ==
     prop2cnf_logic(prop, make_predata(clist, pcnt, ccnt)) &&
     __return == res && res != 0 && res <= pcnt' && -res <= pcnt' &&
     Zlength(clist') <= Zlength(clist) + 4 * SmtProp_size(prop) &&
     pcnt' <= pcnt + SmtProp_size(prop) &&
     store_SmtProp(p, prop) *
     store_predata(data, clist', pcnt', ccnt')
*/    
\end{minted}

\section{Example: multiple specifications and specification derivation}\label{appendix:multispec}

We use the \lstinline{swap} function as an example to illustrate the different approaches of multiple specifications and specification derivation. As is well known, the \lstinline{swap} function has two specifications in verification scenarios: one where the input parameters occupy the same address and another where they occupy different addresses. Using the multiple specifications method, this can be expressed as follows:

\begin{lstlisting}
void swap(int * px, int * py)
//@ requires integer(px, ?x) &*& integer(py, ?y);
//@ ensures integer(px, y) &*& integer(py, x);
//@ requires px == py &*& integer(px, ?x);
//@ ensures px == py &*& integer(px, x);
\end{lstlisting}

Based on specification derivation, we can express it as follows:

\begin{lstlisting}
void swap(int * px, int * py)
/*@ all
    With para
    Require swap_pre(px, py, para)
    Ensure swap_post(px, py, para)
*/;

void swap(int * px, int * py)
/*@ neq <= all
    With x y
    Require x == *px && y == *py
    Ensure  y == *px && x == *py
*/;

void swap(int * px, int * py)
/*@ eq <= all
    With x
    Require px == py && x == *px
    Ensure  x == *py
*/;
\end{lstlisting}

Here, we introduce \lstinline{para} to distinguish between the cases of same address and different addresses. In the actual proof of \lstinline{all} specification, we manually perform case analysis on \lstinline{para} to expand it into the two scenarios, and then complete subsequent symbolic execution and verification condition solving within the multiple branches. In this example, since the different specifications are entirely independent and cannot be derived from one another, using multiple specifications is more convenient. Although we have demonstrated how to achieve a similar effect using specification derivation, QCP will also support multiple specifications in future versions.
\end{document}